\documentstyle[preprint,aps]{revtex}

\begin{document}
\title{Some comments on superconductivity in diborides}
\author{D.Kaczorowski, J.Klamut, A.J.Zaleski}
\address{Institute of Low Temperature and Structure Research, PAS, Wroclaw, Poland}
\maketitle

\begin{abstract}
Short discussion trying to explain, why superconductivity revealed for some
diborides is not always confirmed in experiments of different research
groups.
\end{abstract}

\bigskip The discovery of superconductivity in MgB$_{2}$ by Nagamatsu et al. 
\cite{Nagamatsu} awakened superconductivity community who waited till
satisfactory theory of high temperature superconductivity will be developed.
Although it looks impossible, the phenomenon was discovered in very simple,
two-component compound and its critical temperature was surprisingly high.
The activities during the first month after the presentation of the results
by Akimitsu \cite{Akimitsu} was the proof how well we are prepared for the
study of new superconducting materials.

But MgB$_{2}$ is only a member of a rich family of diborides. So also the
other members of this family (Li, Be, Al, Ti, Zr, Hf, V, Nb, Ta, Cr, Mo)
became the subject of intensive studies of different groups.

Already in 1970 superconductivity was discovered by Cooper et al. \cite
{Cooper} in NbB$_{2}$ with critical temperature equal to T$_{c}$ = 3.87 K
and in Zr$_{0.13}$Mo$_{0.87}$B$_{2}$ with T$_{c}$ above 11 K.

Systematic study of diborides was conducted by Leyarovska et al. \cite
{Leyarovska}. They looked for superconductivity in these compounds at
temperatures down to 0.42 K and showed that only NbB$_{2}$ was
superconducting at T$_{c}$ = 0.62 K. They didn't check MgB$_{2}$. (As an
additional interesting fact it can be added that they also didn't check the
compound UBe$_{13}$, which they had got, but which was beyond the scope of
their study. We can only try to imagine what would be if superconductivity
in heavy fermion UBe$_{13}$ and MgB$_{2}$ had been discovered already in
1979...)

Reexamining of the properties of diborides leaded to interesting results.
Some groups of researchers found superconductivity for compounds for which
other groups found no such an effect.

Kaczorowski et al. \cite{Kaczorowski} found superconducting transition at T$%
_{c}$= 9.5 K for TaB$_{2}$ and no superconductivity for TiB$_{2}$, HfB$_{2}$%
, VB$_{2}$, NbB$_{2}$ or ZrB$_{2}$ . Although Felner \cite{Felner} stated
that BeB$_{2}$ is not superconducting, according to Young et.al \cite{Young}
BeB$_{2.75}$ is superconducting with T$_{c}\approx $ 0.7 K. Gasparov et al. 
\cite{Gasparov} found ZrB$_{2}$ superconducting with T$_{c}$ = 5.5 K and
simultaneously they did not confirm superconductivity for TaB$_{2}$ and NbB$%
_{2}$.

Superconductivity in TaB$_{2}$ was discovered in old, well aged material.
Although from DC magnetization measurements it resulted that practically
100\% of volume of the sample was superconducting, authors tried to prove
that observed diamagnetic signal was not connected with some spurious phase.
The main candidates for such phases were tantalum (or niobium) oxides and
carbonates. But even if some of them have their critical temperatures
similar to that measured in \cite{Kaczorowski}, their upper critical fields
were much lower than measured TaB$_{2}$ being equal to H$_{c2}$(0) = 2.3 T.
The only ''impurity-type'' explanation could be connected with the existence
of some amount of tetragonal $\beta $-Ta , which could form superconducting
phase with boron.

It also should be added that hydrogenation of superconducting TaB$_{2}$
resulted in rather significant hydrogen uptake above 30\% and decrease of
the amount of superconducting phase with the practically unchanged critical
temperature \cite{Zaleski}. This result also suggests that it is not
tantalum oxide which is superconducting in the sample. It is also needless
to say, that all tests to find oxygen or carbon using EDAX have failed.

Attempts to prepare the new material with the same composition i.e. TaB$_{2}$
resulted in obtaining compound which had transition temperature about 10K,
but calculated volume of superconducting material was within few percent.
Similar result of superconducting transition of small percentage of the
sample volume of nominal composition TaB$_{2}$ was also obtained in Dresden 
\cite{Nenkov} It was also found that small changes in relative amount of
tantalum and boron resulted in material possessing quite different magnetic
properties (this issue is under systematic study now). Just these findings
were the reason why the role of sub- and superstoichiometric compositions
range in diborides was emphasized in \cite{Kaczorowski}.

Recently this preassumption was supported by the paper by Young et al. \cite
{Young} where superconductivity was obtained in the material with meaningful
excess of boron.


\begin{references}
\bibitem{Akimitsu}  J.Akimitsu, Symposium on Transition metal Oxides,
Sendai, January 10, 2001

\bibitem{Nagamatsu}  J.Nagamatsu, N.Nakagawa, T.Muramaka, Y.Zenitani,
J.Akimitsu, Nature {\bf 410}, 63 (2001)

\bibitem{Cooper}  A.S.Cooper, E.Corenzerst, L.D.Longinotti, B.T.Matthias,
W.H.Zachariasen, Proc.Natl.Acad.Sci. {\bf 67}, 313 (1970)

\bibitem{Leyarovska}  L.Leyarovska, E.Leyarovski, J.Less-Common Metals, {\bf %
67}, 249 (1979)

\bibitem{Kaczorowski}  D.Kaczorowski, A.J.Zaleski, O.J.\.{Z}oga\l ,
J.Klamut, condmat/0103571 (2001)

\bibitem{Felner}  I.Felner, cond-mat/0102508 (2001)

\bibitem{Young}  D.P.Young, P.W.Adams, J.Y.Chan, F.R.Fronczek,
cond-mat/0104063 (2001)

\bibitem{Gasparov}  V.A.Gasparov, N.S.Sidorov, I.Izver'kova, M.P.Kulakov,
cond-mat/0104323 (2001)

\bibitem{Zaleski}  A.J.Zaleski, W.Iwasieczko, M.Tkacz, D.Kaczorowski,
H.Drulis, O.J.\.{Z}oga\l , J.Klamut - to be published

\bibitem{Nenkov}  K.Nenkov - private communication
\end{references}
\end{document}